\documentclass[pra,twocolumn,footinbib,superscriptaddress]{revtex4}

\usepackage[latin1]{inputenc}
\usepackage{graphicx}
\usepackage{amsmath}

\begin{document}
\title{Influence of molecular temperature on the coherence 
\\of fullerenes in a near-field interferometer}
\author{Klaus Hornberger}
\affiliation{Institut f\"{u}r Experimentalphysik, Universit\"{a}t
  Wien, Boltzmanngasse 5, 1090 Wien, Austria }
\affiliation{Department f{\"u}r Physik, Universit\"{a}t M\"{u}nchen,
Theresienstra{\ss}e 37, 80333 M\"{u}nchen, Germany }
\author{Lucia Hackerm{\"u}ller}
\author{Markus Arndt}
\affiliation{Institut f\"{u}r Experimentalphysik, Universit\"{a}t
  Wien, Boltzmanngasse 5, 1090 Wien, Austria }


\begin{abstract}
  We study C$_{70}$ fullerene matter waves in a Talbot-Lau
  interferometer as a function of their temperature.  While the ideal
  fringe visibility is observed at moderate molecular temperatures, we
  find a gradual degradation of the interference contrast if the
  molecules are heated before entering the interferometer. A method is
  developed to assess the distribution of the micro-canonical
  temperatures of the molecules in free flight.  This way the
  heating-dependent reduction of interference contrast can be compared
  with the predictions of quantum theory. We find that the observed
  loss of coherence agrees quantitatively with the expected
  decoherence rate due to the thermal radiation emitted by the hot
  molecules.
\end{abstract}
\maketitle

\section{Introduction}

The wave-particle duality of material objects is a hallmark of quantum
mechanics.  Up to now, the wave nature of particles has been
demonstrated for electrons, neutrons, atoms, and coherent
atomic ensembles. Recently, even the interference of composite
objects has been observed ranging from molecular
dimers \cite{Chapman1995b,Borde1994a,Schollkopf1996a} and van der Waals
clusters~\cite{Schoellkopf2004a} to fullerenes~\cite{Arndt1999a},
massive fullerene derivatives, and small
biomolecules~\cite{Hackermuller2003a}. In particular, the advances with
large molecules have stimulated the question what determines the
limits to observe quantum delocalization with massive objects.

From the theoretical side, there is also an increased interest
concerning the location of the apparent quantum-classical boundary.
The recent understanding of decoherence phenomena points to the
crucial role played by the environmental interaction in determining
whether a quantum particle shows wave behavior \cite{Joos2003a}.

Several experiments tackled this issue in the context of
interferometry: Pritchard and coworkers studied the loss of
interference contrast in an atom interferometer were sodium atoms were
subjected to resonant laser light~\cite{Chapman1995a,Kokorowski2001a}.
The photons emitted in a spontaneous decay then both imparted a recoil
on the atom and entangled the atomic state with the escaping photon
state. Similarly, the Paris experiment \cite{Brune1996a} can be
regarded as an internal state interferometer, where the fringe
visibility depends on the entanglement between the atom and the field
mode of the emitted photon~\cite{Bertet2001a}.  Decoherence has also
been investigated in a Mach-Zehnder interferometer with a 2D electron
gas in a semiconductor heterostructure~\cite{Ji2003a}. The Heiblum
group observed a loss of fringe visibility when the mesoscopic
structure was heated.

Here, we describe in detail our recent molecule interferometry
experiment which explores the loss of fringe contrast when the
internal degrees of freedom of fullerene molecules are heated to
temperatures above 2000\,K. It is based on the fact that hot
fullerenes emit thermal, visible light in a continuous
spectrum~\cite{Mitzner1995a}. One expects that the emitted photons
will reduce the fringe contrast once they are energetic enough to
perturb the motional state of the molecule, or equivalently, if they
convey sufficient information about its
position~\cite{Hornberger2004a}. While the original observation has
been published in~\cite{Hackermuller2004a}, the present article
provides a detailed account of the experimental and theoretical
techniques used to determine the  temperature of
the fullerenes in free flight which is needed to calculate the
expected decoherence effect.

In Sect.~\ref{sec:setup} we describe the setup of the experiment and
our method for varying the molecular temperature by laser heating.
Section~\ref{sec:cooling} provides a quantitative description of the
thermally emitted radiation, which determines the localization of the
molecular waves and their cooling process. Knowledge of the latter is
required in Sect.~\ref{sec:thermometrymodel} where a model is
presented that quantifies the competing dynamics between ionization
and cooling of the fullerenes.  The comparison with experimental data
in Sect.~\ref{sec:temperatures} allows to extract the temperature
distribution in the beam. In Sect.~\ref{sec:decoherence} we compare
the observed loss of visibility to the expectations from quantum
theory, and we discuss alternative decoherence mechanisms in
Sect.~\ref{sec:conclusions}.

\section{Interferometry with heated fullerenes}

\label{sec:setup}

\subsection{The interferometer}
\label{sec:interferometer}

\begin{figure}
[ptb]
\includegraphics[width=\columnwidth]
{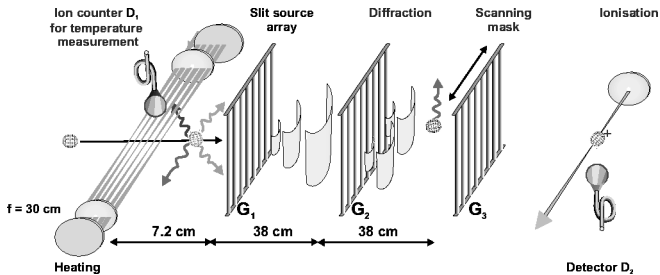}
\caption{The Talbot-Lau-Interferometer consists of three equal
  gratings with a slit separation of $d=0.99\,\mu$m spaced by a
  distance of $L=38$\,cm. The interference contrast can be reduced by
  heating the fullerenes in front of the interferometer via multiple
  laser beams. }
\label{setup}
\end{figure}

We use a Talbot-Lau matter wave interferometer as already described in
detail in~\cite{Brezger2002a}. It consists of three gold gratings with
a period of $d=990$\,nm and a nominal open width of 470\,nm which are
equally separated by the distance $L=38$\,cm, see Fig.~\ref{setup}.
The first grating of the setup prepares the wave coherence in the
beam. Near-field diffraction at the second grating then produces a
high-contrast interference pattern at the position of the third
grating provided the grating separation $L$ is close to a multiple of
the Talbot length $L_{\rm T}={d^2}/{\lambda_{\rm C70}}$
\cite{Clauser1994a}.  At a grating separation of $L=$0.38\,m the
interference fringes corresponding to the first and the second Talbot
order are expected at molecular de Broglie wavelengths of
$\lambda_{\rm C70}=2.6$\,pm and of $\lambda_{\rm C70}=5.2$\,pm,
respectively.  Different height constrictions are used to select the
required velocities centered at 100\,m/s ($\Delta v/v\simeq 10\%$) and
at 190\,m/s ($\Delta v/v \simeq 15\%$) out of a thermal beam of
fullerenes, which is produced by sublimation at a temperature of
900\,K. The interference pattern has a period equal to the grating
constant $d$. Therefore one observes modulation fringes in the
transmitted flux $I$ if the third grating is shifted laterally.  The
transmitted molecules are detected behind the third grating by laser
ionization.  The observed contrast of the fringe signal and its
wavelength dependence then prove unambiguously the coherence of the
interference effect.

\begin{figure}
[ptb]
\includegraphics[width=\columnwidth]
{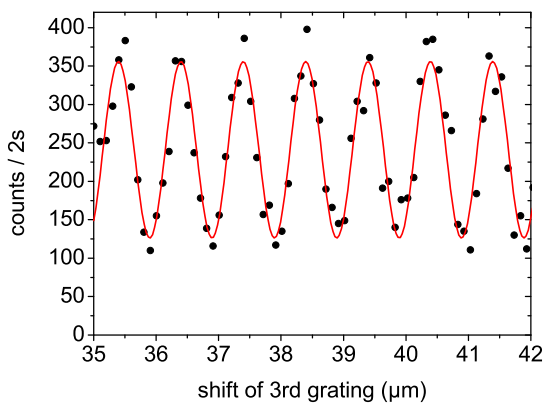}
\caption{High contrast Talbot-Lau interference fringes are observed
for fullerenes at temperatures below 1000\,K. The fringe visibility of
47\% corresponds to the quantum mechanical expectation indicating 
coherent molecular evolution.}
\label{fringes}
\end{figure}

As shown in earlier work~\cite{Brezger2002a,Hackermuller2003a} the
Talbot-Lau setup is well suited for investigating the wave nature of
large molecules such as fullerenes, tetraphenylporphyrins and
fluorofullerenes. As an example, Fig.~\ref{fringes} gives the
interference signal of C$_{70}$ fullerenes as observed without
additional heating. The sinusoidal fringes show the expected
visibility $V=(I_{\rm max}-I_{\rm min})/(I_{\rm max}+I_{\rm min})$ of
47\%.

\subsection{The heating stage}
\label{sec:heatingstage}

This interferometric setup is now complemented by a heating stage
about 1 m behind the oven and 7 cm in front of the first grating. It
serves to vary the internal energy of the molecules by photon
absorption~\cite{Hackermuller2003a}. The green fraction of an Argon
ion laser ($\lambda=514$\,nm) is used for the heating stage, while the
blue line ($\lambda=488$\,nm) serves the post-ionization in the detector. The
green line can reach a maximal power of $P_{0}=10.8$\,W and is focused
by a lens of 30\,cm focal length to a waist of $w_{y} =50\mu$m. The
beam passes a cat-eye arrangement of two lenses and two mirrors under
small angles and can thus interact up to 16 times with the molecular
beam, as shown in Fig.~\ref{setup}. The distance between the foci of
the laser beams is approximately 0.3 mm on average. The beam undergoes
some attenuation due to some absorption and reflection in the optical
elements. The power after the $N^{\mathrm{th}}$ reflection is measured
to be about $P_{N}=(11.2-0.42\,N)$\,W. The power of the blue detecting
laser beam is fixed to 16\,W and focused to a waist of
$w_{y}=8\,\mu$m.

The overlap of the laser beam with the molecular beam is monitored by
the count rate at detector D$_2$, which shows a maximum for an optimal
overlap.  All heating beams are then adjusted one after the other with
respect to maximal increase in the count rate.  Molecules which are
heated in front of the interferometer maintain much of their internal
energy until they reach the detector behind the interferometer. Since
the latter is based on thermal ionization \cite{Ding1994a,Nairz2001a}
an increase of the internal molecular energy leads to an enhanced
ionization efficiency in the final detection stage.  With a
molecular beam height of 150 $\mu$m and a heating laser beam waist of 50
$\mu$m we are sure that the laser beams overlaps 
with the molecular beam as long as we can detect an increase for each
additional laser beam.

\subsection{Laser heating of fullerenes}
\label{sec:heatingprocess}

Fullerenes are particularly well-suited for studying temperature
effects in molecular interference since their cage structure provides
them with an extraordinary stability against fragmentation. This
allows us to deposit more than 100\,eV in a single molecule for a
sufficiently long time.  Laser excitation of these molecules has been
studied extensively in the literature, see~\cite{Dresselhaus1998a} and
references therein. Compared to the more symmetric C$_{60}$ molecule
there are fewer symmetry restrictions for dipole transitions in
C$_{70}$. As a result the absorption cross section of C$_{70}$ exceeds
that of C$_{60}$ by almost an order of magnitude. This is the main reason for
our preference of the C$_{70}$ molecule.

\begin{figure}
[ptb]
\includegraphics[width=0.6\columnwidth]
{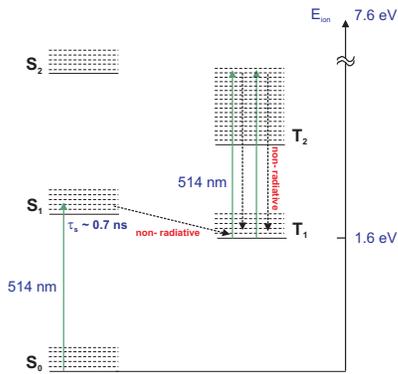}
\caption{Electronic excitation energies of C$_{70}$
  \cite{Dresselhaus1998a}. The first photon excites the S$_{1}$ state
  followed by a rapid non-radiative conversion to the relatively
  long-lived triplet state T$_{1}$. All further photon absorption
  takes place in the electronic triplet configuration.}
\label{levelscheme}
\end{figure}

According to Ref.~\cite{Dresselhaus1998a} the absorption of a 514\,nm
photon from the electronic ground state S$_0$ will excite the singlet
state S$_1$.  This is followed by a rapid non-radiative transition to
the metastable triplet state T$_1$ (the branching ratio exceeds 90\%).
The relevant energy levels and transitions are summarized in
Fig~\ref{levelscheme}.  Given the relatively long life time of T$_1$
and the short life times of the other triplet states it is plausible
that all further excitation after the first photon occurs sequentially
from T$_1$. The corresponding photon energies absorbed from the
heating beam are rapidly transferred to the 204 vibrational degrees of
freedom of the molecule.  The molecules take about 0.4\,ms after the
last heating beam until they enter the interferometer.

For fullerenes there are several processes that lead to a loss of the
energy stored in the vibrational degrees of freedom.  The first
relevant mechanism, which is dominant at large internal energies, is
thermal emission of radiation. It was observed by several groups that
laser heated fullerenes emit visible light in a continuous spectrum
\cite{Rohlfing1988a,Mitzner1995a,Heszler1997a}. The form of this
spectrum and its dependence on the internal energy determines both the
cooling and the decoherence of the molecules, and it is discussed in
detail in the next section.

An important competing process is the thermionic emission of
electrons. This effect was  observed in Ref. \cite{Ding1994a} and
forms the basis of our detection method \cite{Nairz2000a}.  Moreover,
the thermal ions in the heating stage also provide important
information for the assessment of the stored energy of the molecules.

A third thermally activated process, the emission of carbon dimers,
can be safely disregarded in our experiment since the ground state
fragmentation energy is 10.6\,eV ~\cite{Matt1999a}, while the
corresponding ionization potential is 7.6\,eV. This energy difference
is considerably larger than the maximal energy per vibrational mode in
our experiment of about 0.5\,eV.  As expected, we do not observe any
fullerene fragments such as C$_{68}$ in a quadrupole mass spectrometer
mounted behind the interferometer.

By varying the power of the heating laser beam, our setup allows us to
control the internal energy of the fullerenes stored in the
vibrational motion of the carbon nuclei.  In order to quantify the
decoherence caused by the corresponding heat radiation a detailed
understanding of the radiative cooling process is indispensable. It
is developed in the next section.  After that, in
Sect.~\ref{sec:thermometry}, we show how to describe the competing
dynamics between the cooling process and the thermal
ionization. 

\section{Radiative cooling}
\label{sec:cooling}

The 204 vibrational modes of C$_{70}$ fullerenes form a heat reservoir
that can store large amounts of energy. Irrespective of the type of
heating, it is sufficient for our purposes to characterize the
vibrational state of the molecule by the total internal energy $E$.
Moreover, it is convenient to specify $E$ in terms of a temperature,
even though the molecules are insulated from any external heat bath.
This can be done by the micro-canonical temperature $T^{}(E)=\left[
  \partial S/\partial E\right] ^{-1}$, which is defined through the
entropy $S(E)$.  To leading order, $T^{}$ is the temperature of a
fictitious heat bath that would be required to keep the average
internal energy of the particle at a value of $E$.

In free flight the loss of internal energy, i.e., the cooling of
the neutral molecules is solely determined by the emission of thermal
photons.  Ultimately, this process determines the decoherence of the
molecular wave since the thermal photons may carry which-path
information.

\subsection{The fullerene emission spectrum}

The thermal radiation emitted from fullerenes was observed to have a
continuous spectrum \cite{Mitzner1995a}. However, the spectral
emission rate deviates from the Planck law of a macroscopic black body
for a number of reasons. First, the radiating particle is much smaller
than the typical photon wavelengths which indicates that it must be a
colored emitter. The oscillator strengths of the available transitions
can be related to the frequency dependent absorption cross section
\cite{Friedrich1998a}. Second, at the internal energies where thermal
emission is relevant the particle is not in thermal equilibrium with
the ambient radiation field so that there is no induced emission.
Third, the emitter is not kept at a constant temperature, but the
emission takes place at a fixed energy $E$.  Similar to Einstein's
derivation of the Planck law these aspects lead to the expression
\cite{Hansen1998a}

\begin{align}
R_{\omega}\left(  \omega,T^{}\right)  \mathrm{d}\omega=&\frac{\omega^{2}
}{\pi^{2}c^{2}}\sigma_{\mathrm{abs}}\left(  E(T^{})-\hbar\omega
;\omega\right)  
\label{Romega}
\\
&\times
\exp\left[  -\frac{\hbar\omega}{k_{\mathrm{B}}T^{}}
-\frac{k_{\mathrm{B}}}{2C_{V}}
\left(  \frac{\hbar\omega}{k_{\mathrm{B}}T^{}}\right)
^{2}\right]  \mathrm{d}\omega\,.\nonumber 
\end{align}
The first term is proportional to the mode density. The second term,
the absorption cross section at internal energy $E(T^{})-\hbar\omega$
and frequency $\omega$ quantifies the strength of the available
electronic transitions.  We disregard the temperature dependence of
$\sigma_{\mathrm{abs}}$ since the relevant optical transitions have
much higher energies than the thermal excitations. The third term in
(\ref{Romega}), the statistical factor, contains the first-order
correction due to the finite heat capacity $C_V$ of the fullerenes
\cite{Hansen1998a,Hornberger2004a}. It takes into account that the
photon emission occurs at a fixed total energy rather than at a fixed
temperature.

The heat capacity of C$_{70}$ is practically constant at the
temperatures between $1000\,\mathrm{K}$ and $3000$\,K
\cite{Kolodney1995a}. It was calculated from the energies of the
vibrational levels to be about $C_{V}=202$ $k_{\mathrm{B}}$.

\begin{figure}
[ptb]
\includegraphics[width=\columnwidth]
{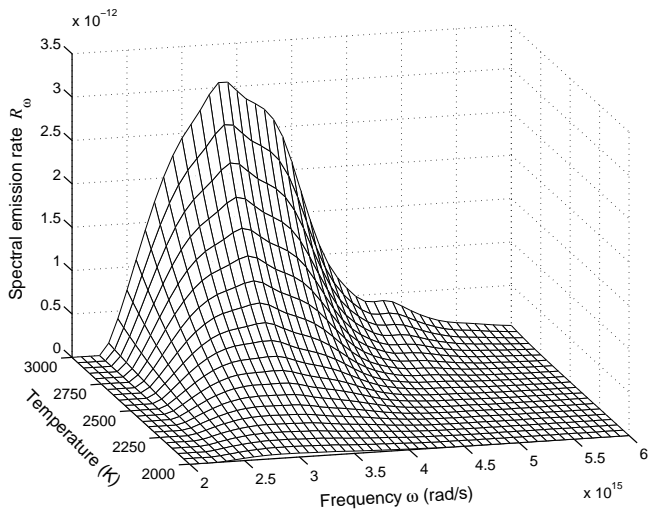}
\caption{The calculated spectral photon emission rate for
  (micro-canonical) molecular temperatures between 2000\,K and 3000\,K,
  cf.~Eq.~(\ref{Romega}). It involves a measured frequency dependent
  absorption cross section \cite{Coheur1996a}. The data imply that
  about three visible photons are emitted during a transit time of
  4\,ms.  }
\label{emissionrate}
\end{figure}

Using measured data for the absorption cross
section~\cite{Coheur1996a} we obtain the spectral emission rate shown
in Fig.~\ref{emissionrate}. It is given for the temperature range
between 2000\,K and 3000\,K that is relevant in our experiment and
covers the range of visible frequencies. Note that there are no
contributions in the infrared because of a lack of accessible
transitions below 1.6 eV, the energy of the HOMO-LUMO gap in the
molecule \cite{Dresselhaus1998a}. Although some of the 204 vibrational
modes of C$_{70}$ are radiative in the far infrared
\cite{Dresselhaus1998a}, they do not contribute appreciably to
decoherence or cooling in our experiment since the corresponding
transition matrix elements are extremely small \cite{Andersen2000a}.
The small bump that can be seen at large temperatures around
$\omega=4.8\times 10^{15}$\,rad/s corresponds to the energy of
3.16\,eV belonging to the first allowed dipole transition.  The other
direct dipole transitions and the plasma resonance at greater energies
are still suppressed below 3000\,K by the Boltzmann factor in
(\ref{Romega}).

\subsection{The cooling dynamics}

The radiative cooling of the fullerene molecules is determined by the
spectral emission rate through the total radiant flux
\begin{equation}
\Phi\left( T\right) =\int_0^\infty\hbar\omega R_{\omega}\left(
\omega,T\right) \mathrm{d}\omega. \label{Phi}
\end{equation}
It is important to note that the temperature dependence of $\Phi$
differs considerably from the Stefan-Boltzmann law in the case of
fullerenes, due to the gap in the electronic excitation spectrum
mentioned above. Indeed, from Eqs. (\ref{Romega}) and (\ref{Phi}) one
finds that the radiant flux of C$_{70}$ is  given
approximately by $\Phi\left( T\right) =6.3\times10^{-35}
(T/\mathrm{K})^{11}\,\mathrm{eV/s}$ in the temperature regime of
$T=2000~\mathrm{K}-3000$ K which will be relevant below. It shows that
the thermal emission of fullerenes increases much more strongly with
temperature than the well-known $T^{4}$-dependence of a black body.

It follows that the temporal evolution of a hot, neutral fullerene is
governed by the cooling equation

\begin{equation}
\frac{\mathrm{d}}{\mathrm{d}t}T(  t)
=-\frac{\Phi\left(  T\right)  }{C_{V}} \label{Tdeq},
\end{equation}
By integrating (\ref{Tdeq}) numerically for a finite time $t>0$ we
find that the reduction of an initial temperature $T_{0}$ is 
described extremely well by the approximate formula
\begin{equation}
T\left(  t;T_{0}\right)  =T_{0}\left(  1+\left(  \frac{T_{0}}{T_{\infty}
}\right)  ^{n}\right)  ^{-1/n}, 
\label{Tdecay}
\end{equation}
with the time dependent parameters $n$ and $T_{\infty}.$ From this it follows
that an initial temperature distribution $f_{0}\left(  T\right)  $ transforms
as
\begin{align}
f_{t}\left(  T\right)  =&f_{0}\left(  T\left(  1-\left(  \frac{T}{T_{\infty}
}\right)  ^{n}\right)  ^{-1/n}\right)
\nonumber\\
&\times
  \left(  1-\left(  \frac{T}{T_{\infty}
}\right)  ^{n}\right)  ^{-(n+1)/n}. 
\label{ftransform}
\end{align}
Note that any temperature distribution will be bounded from above
after a finite time by the maximum temperature $T_{\infty}$, as
implied by Eq.~(\ref{Tdecay}).

The parameters $n$ and $T_{\infty}$ are obtained by fitting the
function (\ref{Tdecay}) to the numerical solution of Eq.~(\ref{Tdeq}).
For the short times of flight $t$ between the heating beams we find
$n=6.2,$ $T_{\infty}=830\,\mathrm{K} \,\left( \mathrm{s}/t\right)
^{1/n}.$ The longer passage from the heating beams to the first
grating of the interferometer is characterized by $n=8.5$ and
$T_{\infty}=1700\,\mathrm{K} \,\left(v\,\mathrm{s/m}\right) ^{1/n},$
with $v$ the velocity.  During the long flight from the first grating
to the detection laser the cooling is described by $n=10.5,$
$T_{\infty }=2166\,\mathrm{K}$ for a fullerene velocity of
100\thinspace m/sec, and by $n=9.7,$ $T_{\infty}=2321\,\mathrm{K}$ for
190\thinspace m/s, respectively.  Finally, we have $n=9,$
$T_{\infty}=1490\,\mathrm{K}\,\left( v\,\mathrm{s}/\mathrm{m}\right)
^{1/n}$ for the range of about 30\,cm behind the detection laser where
ions are recorded.

\section{Molecular thermometry: Theoretical description}
\label{sec:thermometry}
\label{sec:thermometrymodel}

After the fullerene beam has passed the heating stage it is no longer
characterized by a single internal temperature.  Rather, it shows a
distribution of temperatures because of the stochastic nature of the
photon absorption and as a result of the different heating intensities
seen by the fullerenes due to the laser beam profiles.  The resulting
average temperature cannot be measured directly with a spectrometer
since there are too few detectable photons emitted while the
fullerenes travel through the interferometer.  Instead, we have to
resort to an indirect assessment of the temperature distribution.

The beam temperature is inferred in our experiment from an accurate
measurement of the molecular ions, which appear due to the thermal
emission of electrons from the hot fullerenes in the heating stage.
They are recorded as a function of the fullerene velocity, the laser
power, and the number of heating laser beams.  By comparing a model
calculation to many experimental curves we obtain reliable information
on the temperature distribution in the beam.  A second, independent
source of information is the count rate variation in the final
molecule detector. This is because the fullerenes carry much of their
added energy through the whole interferometer and their temperature
determines the final detection efficiency. As discussed below, our
model calculation also reproduces the experimental detection rate for
different heating powers and fullerene velocity, which confirms
independently the implied temperature distribution.

A quantitative description of the heating dynamics and the resulting
ion signal faces several challenges: First, the photon absorption is a
stochastic process giving rise to an initially broad distribution of
temperatures in the beam.  Second, there is a delicate competition
between radiative emission and ionization, which both depend in a
strong and nonlinear fashion on the fullerene temperature.  Third, the
radiative cooling of the molecules between the subsequent heating
laser beams must also be taken into account.  Finally, the description
is further complicated by the fact that the fullerenes experience
different heating intensities due to the finite extension of the
molecular and optical beams.  In the following model all those aspects
are taken into account, which is necessary to provide a realistic
description of the experimental situation.

We represent the initial velocity distribution by the
function~\cite{Scoles1988a}
\begin{equation}
I_{\mathrm{C70}}\left(  v\right)
~\mathrm{d}v=C_{T}~v^{3}\exp\left(
-\frac{v^{2}}{v_{w}^{2}}\right)  \mathrm{d}v,
\end{equation}
corresponding to an effusive molecular beam.  Here
$v_{w}=\sqrt{2k_{\mathrm{B}}T/m}=133$ m/s is the most probable speed
of the C$_{70}$ molecules in the oven and $C_{T}$ a constant
determined by the oven aperture.  We then introduce the non-ionized
fraction $i_{\mathrm{C70}}\left( v,z\right)$ of the beam at
longitudinal position $z$ and fullerene velocity $v$.  Moreover, we
need the distribution $i_{\mathrm{C70}}\left( T,y;v,z\right)$ of the
non-ionized fraction with respect to the molecular temperature $T$,
and the vertical position $y$. Initially, the distribution is peaked
in $T$ at 900\,K, flat in $y$, and normalized, $\int
\mathrm{d}y\int\mathrm{d}T~i_{\mathrm{C70}}\left( T,y;v,0\right) =1.$
In terms of this quantity the current of neutral fullerenes at
longitudinal position $z$ is given by
\begin{equation}
I_{\mathrm{C70}}\left(  v;z\right)  =C_{T}~v^{3}\exp\left(  -\frac{v^{2}
}{v_{w}^{2}}\right)
\int\mathrm{d}y\int_0^\infty\mathrm{d}T~i_{\mathrm{C70}}\left(
T,y;v,z\right)  .
\end{equation}
The evolution of the distribution of the neutral fraction
$i_{\mathrm{C70}}$ contains all the information needed to extract the
observable ion yield and detection efficiency. We calculate the
dynamics of $i_{\mathrm{C70}}$ by sequences of transformations which
describe the heating experienced by crossing a single laser beam and
the subsequent cooling and ionization.

In doing so we allow for an absorption cross section $\sigma
_{\mathrm{abs}}$ in the metastable T$_1$ state which may differ from
the ground state absorption cross section. At the same time only
sequential absorption is considered, since the neglect of multi-photon
effects is certainly justified at the prevailing laser intensities.

The absorption of laser photons is determined by a poissonian
probability distribution. Every green photon adds 2.3\,eV to the
molecule and therefore increases the temperature by $\Delta
T=\hbar\omega_{\mathrm{L}}/C_{V}=139$\,K. Correspondingly, each
passage through a single laser beam changes the distribution in the
fullerene beam according to
\begin{align}
i_{\mathrm{C70}}^{\prime}\left(  T,y;v,z_{\mathrm{L}}\right)  =&\mathrm{e}
^{-\bar{n}\left(  v,y\right)  }\sum_{n=0}^{\infty}\frac{\left[
\bar{n}\left( v,y\right)  \right]
^{n}}{n!}
\nonumber\\&\times 
i_{\mathrm{C70}}\left(T-n\Delta T,y;v,z_{\mathrm{L}}\right)  . 
\label{heattrafo}
\end{align}
The mean number of absorbed photons follows from an integration
over the
gaussian laser mode centered at $y_{0}.$ It is given by
\begin{equation}
\bar{n}\left(  v,y\right)
=\sqrt{\frac{2}{\pi}}\frac{\lambda\sigma
_{\mathrm{abs}}}{hcw_{y}v}P\exp\left(  -\frac{2(y-y_{0})^{2}}{w_{y}^{2}
}\right)  .
\end{equation}
The cooling and ionization processes after the heating are governed by
Eq. (\ref{Tdecay}) and by the Arrhenius law \footnote{In the
  literature there are also proposals for a slightly different form of
  the Arrhenius law for fullerenes \cite{Klots1991a}.
  We checked that the result of our calculation described
  below does not depend significantly on the precise temperature
  dependence of the ionization rate.}
\begin{equation}
\frac{\mathrm{d}}{\mathrm{d}z}i_{\mathrm{C70}}\left(
T,y;v,z\right)
=-\frac{A_{\mathrm{ion}}}{v}\exp\left(  -\frac{E_{\mathrm{ion}}}
{k_{\mathrm{B}}T}\right)  i_{\mathrm{C70}}\left(  T,y;v,z\right)
.
\label{dzion}
\end{equation}
In our case the ionization energy is to be taken from the triplet
state T$_{1}$, that is $E_{\mathrm{ion}}=7.6~\mathrm{eV}-1.6\
\mathrm{eV}.$ Since the evolution of the temperature $T(z/v;T_0)$ is
known explicitly (\ref{Tdecay}) one can integrate Eq.
(\ref{dzion}). This way we obtain
\begin{widetext}
\begin{align}
i_{\mathrm{C70}}\left(  T,y;v,z+\Delta z\right)   
=&i_{\mathrm{C70}}\left(T\left(
1-\frac{T^{n}}{T_{\infty}^{n}}\right)  ^{-1/n},y;v,z\right)
\left(
1-\frac{T^{n}}{T_{\infty}^{n}}\right)  ^{-\frac{n+1}{n}}
\exp\left[  -\frac{A_{\mathrm{ion}}L}{v}C\left(
v,T\left( 1-\frac{T^{n}}{T_{\infty}^{n}}\right)  ^{-1/n}\right)
\right]
\label{ioncooltrafo}
\end{align}
with
\begin{equation}
C\left(  v,T_{0}\right)  =n\frac{T_{\infty}^{n}\left(  v\right)
}{T_{\mathrm{ion}}^{n}}\left[  \Gamma\left(  n,\frac{T_{\mathrm{ion}}}{T_{0}
}\right)  -\Gamma\left(  n,\frac{T_{\mathrm{ion}}}{T_{0}}\left(
1+\frac {T_{0}^{n}}{T_{\infty}^{n}\left(  v\right)  }\right)
^{1/n}\right)  \right] .
\end{equation}
Here $\Gamma\left( n,x\right) $ is the incomplete Gamma function and
$k_{\mathrm{B}}T_{\mathrm{ion}}=E_{\mathrm{ion}}.$ Equation
(\ref{ioncooltrafo}) describes the reduction of the neutral fraction
and its temperature variation during a free flight under the joint
action of radiative cooling and ionization.

The transformations (\ref{heattrafo}) and (\ref{ioncooltrafo}) are
applied as often as there are intersecting laser beams, which have an
average separation of 0.3\,mm. We take their vertical positions to be
uniformly distributed within a width of about $1.5w_{y}$, which is the
estimate obtained from the experimental beam alignment.  Our
values for $T_{\infty}\left( v\right)$ and $n$ are given above in
Sect.~\ref{sec:cooling}.

The ion yield in the heating region, which is measured at
detector D$_{1},$ is
given by the change in the current of neutral molecules. It reads

\begin{align}
I\left(  v\right)  =C_{T}~v^{3}\exp\left(
-\frac{v^{2}}{v_{w}^{2}}\right)
\int\mathrm{d}y\int_0^\infty\mathrm{d}T\,
\left(  i_{\mathrm{C70}}\left(T,y;v,0\right)  
-i_{\mathrm{C70}}\left(  T,y;v,z_{\mathrm{G1}}\right)
\right)  .
\end{align}
\end{widetext}
Here we account also for possible delayed ionization by extending the
measurement interval to the interferometer entrance $z_{\mathrm{G1}}.$
However, delayed ionization does not play an important role for the
parameters given below, and virtually no ionization is predicted for
$z>z_{\mathrm{G}1}$.

The final detection stage D$_{2}$ also operates by laser induced
thermionic emission. Therefore, one can calculate its temperature- and
velocity-dependent detection efficiency using Eqs.~(\ref{heattrafo})
and (\ref{ioncooltrafo}) as well.  We use the values given in
Sect.~\ref{sec:heatingstage} for the wavelength of the detection
laser and its waist, and take the maximal ionization distance to be 30\,cm.

\begin{figure*}[ptb]
\includegraphics[width=0.49\textwidth]
{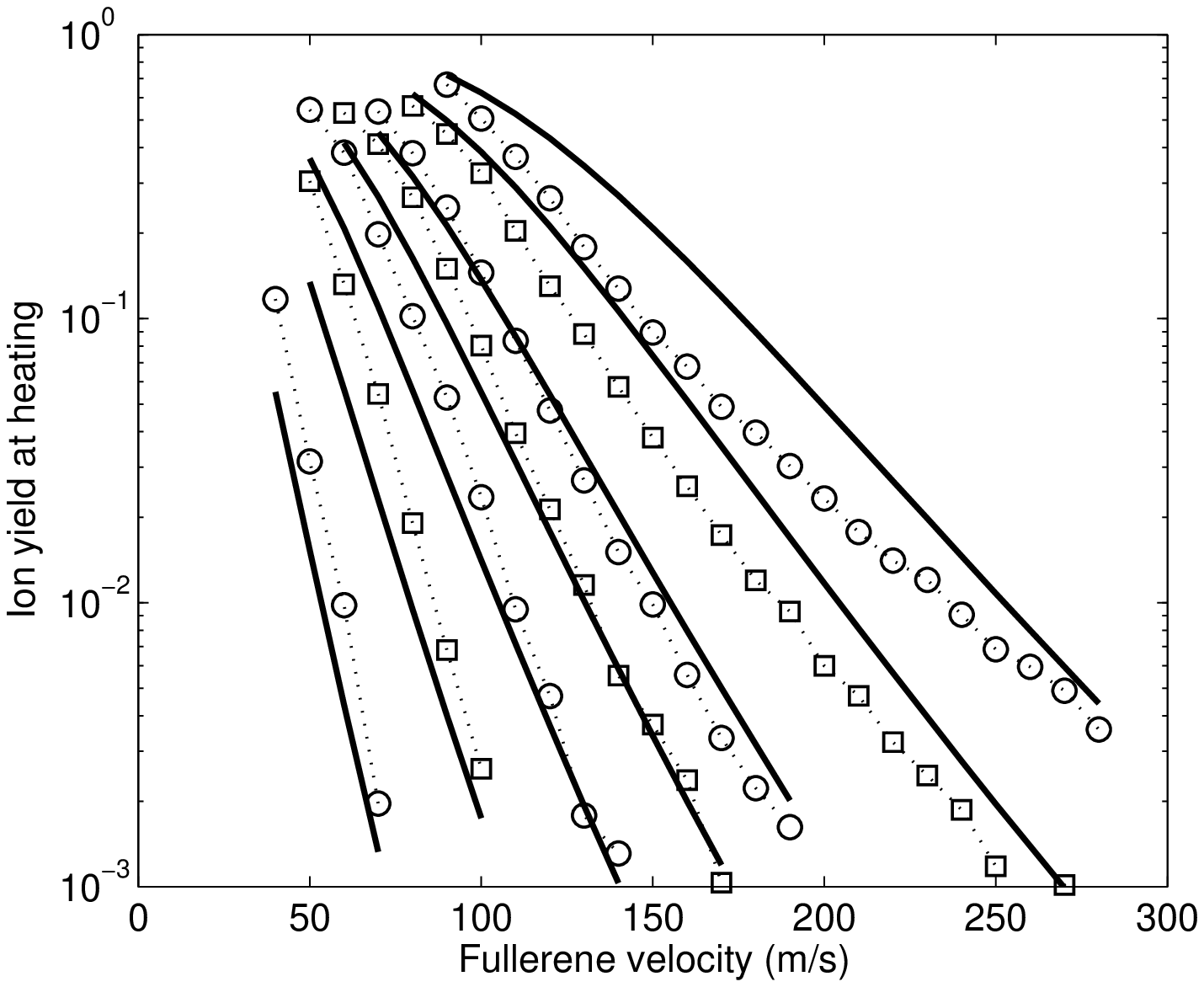}
\includegraphics[width=0.49\textwidth]
{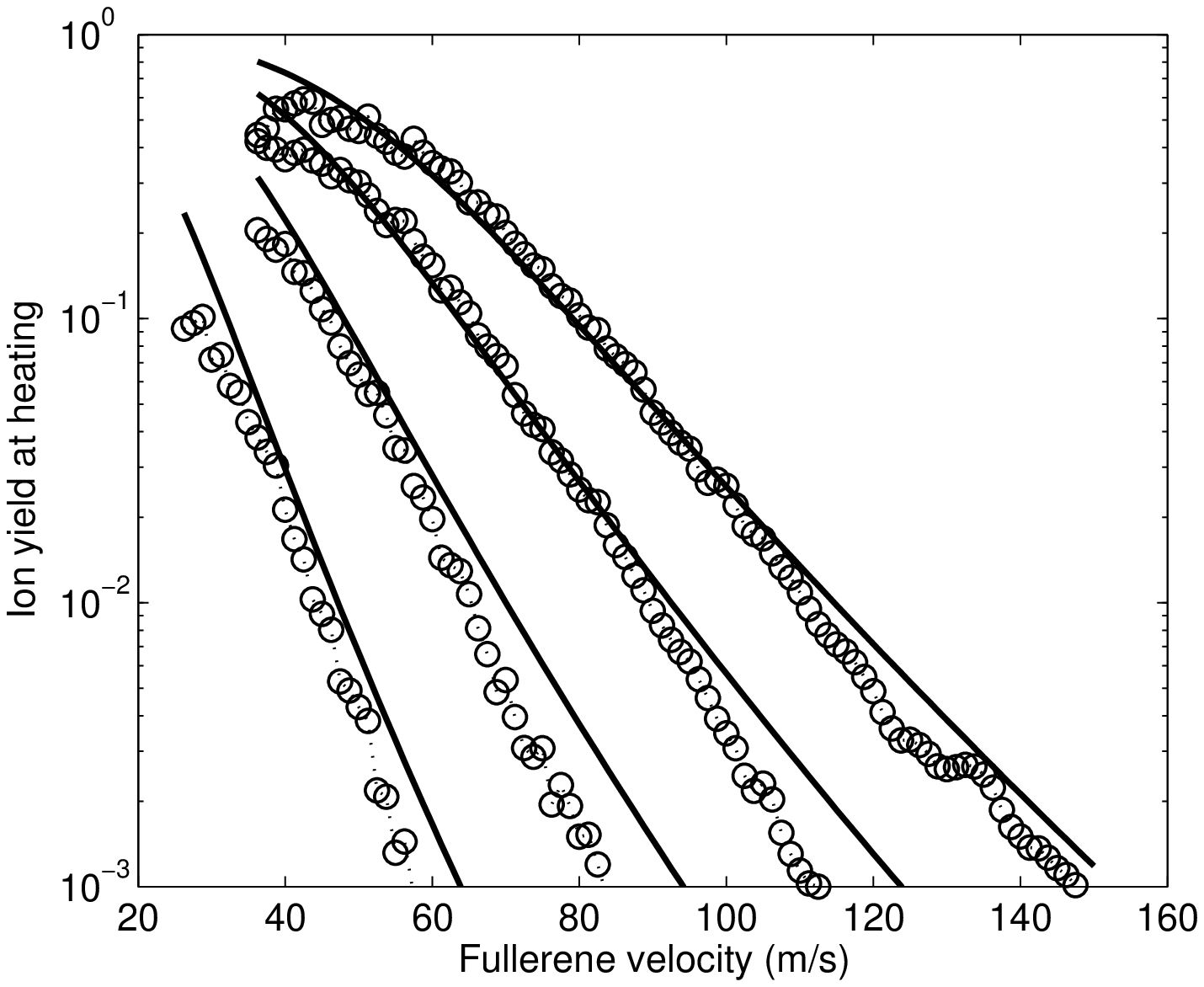}
\caption{Ion yield at the heating stage,
  experimental data (symbols) compared to the model calculation (solid
  lines).  Left: Arrangement with ten heating beams and laser powers
  of $P=2, 3, 4, 5, 6, 8, 10$\,W. Curves with higher count rates are
  associated with higher laser powers. Right: Arrangement with four
  heating beams and laser powers of $P=4, 6, 8, 10$\,W.  }
\label{ions}
\end{figure*}

\section{Molecular thermometry: Experimental results}
\label{sec:temperatures}

In order to gauge our heating setup we measure the ionization yield in
the heating stage.  Positive ions are accelerated to the conversion
dynode where they release secondary electrons which are in turn
accelerated to a secondary electron multiplier situated 1\,cm above
the conversion plate.  The first heating beam was adjusted such that
its focus is lying just below the rim of the Channeltron.  This setup
ensures that the observed electric signal is essentially proportional
to the ion current produced in the heating region.

We use a chopped molecular beam and a time-of-flight sensitive
detection scheme to measure the velocity of the neutral fullerenes,
which determines the heating time.  In addition, the number of heating
beams can be varied by blocking the laser beam after a specified
number of reflections.  This way the normalized ion yield
$I(v)/[C_T v^3 \exp(-v^2/v^2_w)]$ could be extracted 
as a function of the fullerene
velocity, the heating power, and for heating configurations with four
and with ten crossing beams; see Fig.~\ref{ions}.

The corresponding model calculation depends on two unknown parameters,
the absorption cross section in the triplet state $\sigma\left(
\mathrm{T}_{1}\right) $ and the Arrhenius constant for ionization
$A_{\text{ion}}$. By comparing simultaneously the various measured ion
curves to the model we extract the common fit values $\sigma\left(
\mathrm{T} _{1}\right) \simeq2\times10^{-17}$\,cm$^{2}$ and
$A_{\text{ion} }\simeq5\times10^{9}~$s$^{-1}$.  Our value for the
cross section $\sigma\left( \mathrm{T} _{1}\right)$ is close to that
of the known ground state absorption cross section \cite{Coheur1996a}.
This is reasonable with C$_{70}$, where the dipole matrix elements
have fewer symmetry restrictions than C$_{60}$. We did not find
published Arrhenius constants for C$_{70}$, but we note that the
literature values for C$_{60}$ vary by many orders of magnitude
\cite{Reinkoster2001a}.  Given this it seems acceptable that our value
is smaller than the range of literature values for C$_{60}$.
Figure~\ref{ions} shows the experimental data for the ion yield along
with the results of the model calculation using the mentioned
parameters.

\begin{figure*}[ptb]
\includegraphics[width=0.49\textwidth]{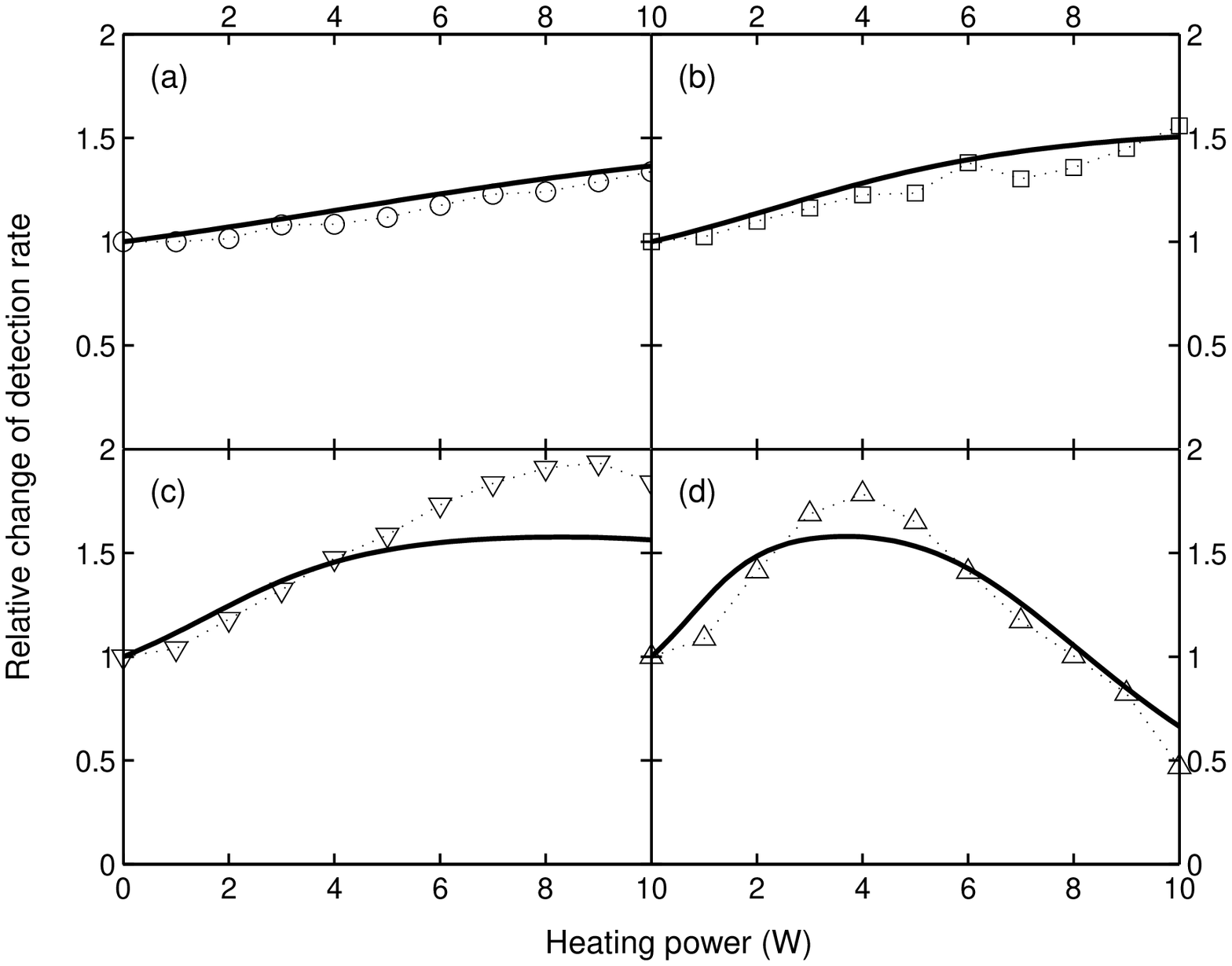}
\includegraphics[width=0.49\textwidth]{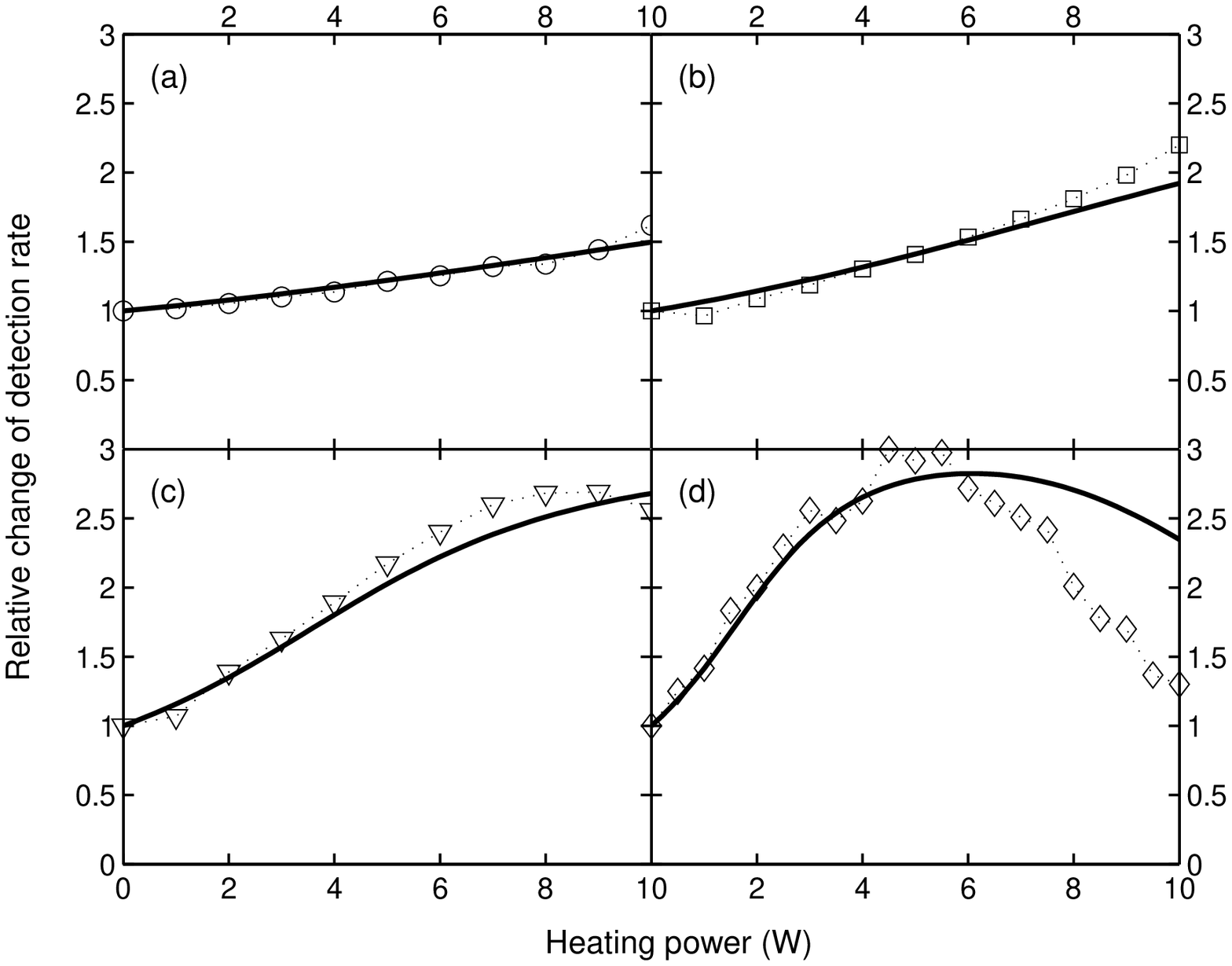}
\caption{Relative change of the molecular detection rate at D$_2$ as a
  function of the heating laser power. Experimental data (symbols) vs.
  model calculation (solid lines). Left: Fullerenes with 100\,m/s. The
  number of heating beams was (a) 1, (b) 2, (c) 4, and (d) 10.  Right:
  Faster fullerenes with a velocity of 190 m/s and (a) 2, (b) 4, (c)
  10, and (d) 16 heating beams.  }
\label{detectorcurves}
\end{figure*}

In addition to the above measurements, we also record the change of
the count rate in the final detection stage as a function of the
heating power, the number of heating beams, and for two different
fullerene velocities of 100\,m/s and 190\,m/s. The experimental data
are shown in Fig.~\ref{detectorcurves}.  They are compared to the
model calculation using the \emph{same} parameters as above, and one
observes a good agreement of the various curves.  This is strong
evidence that the model succeeds in describing the relevant cooling and
ionization processes.  It serves also to explain the  form of the
experimental data curves. The increase in the detection rate at small heating
powers is due to the higher temperature of the fullerenes arriving at
the detection stage.  The increased internal energy raises the
ionization probability after crossing the detection laser beam.  This
effect is superseded at higher heating powers and a large number of
heating beams by a reduction of the detection rate. It is due to the
fact that at strong heating a sizeable fraction of the fullerene beam
is ionized already in the heating stage.

\begin{figure}
[ptb]
\includegraphics[width=\columnwidth]
{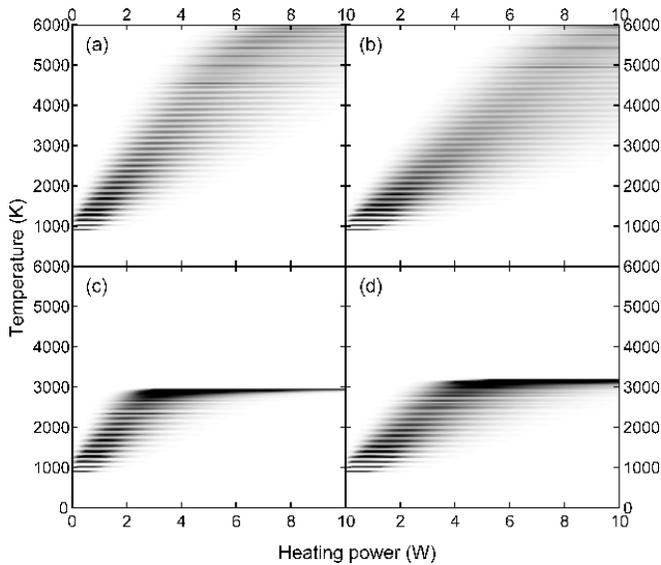}
\caption{Temperature distributions in the beam implied by our model
calculation directly after the heating stage (a),(b) and at the position
of the first grating (c),(d). The shading is proportional to the
probability density of the temperature. The data are shown for a
fullerene velocity of 100 m/s (a),(c) and of 190 m/s (b,d),
respectively. The stripes in (a) and (b) correspond to the temperature
increase $\Delta T=139$\,K due to a single photon absorption.  The
stripes are somewhat closer in (c) and (d) due to cooling.}
\label{tempdistr}
\end{figure}

The good agreement of the described measurements with the model
calculation leads us to conclude that the model provides a realistic
description of the temperature evolution in the beam.
Figure~\ref{tempdistr} shows the implied temperature distribution in
the beam directly after heating and at the position of the first
grating. The fullerene velocities are taken to be 100 m/s and 190 m/s
corresponding to the experimental choice for observing the first and
second Talbot orders. Note that the slower molecules, which are heated
more strongly, arrive at the first grating with a smaller temperature
on average because they have more time of flight to cool off.

A final remark concerns the dependence of the model calculation on the
fit parameters $\sigma\left( \mathrm{T}_{1}\right) $ and
$A_{\text{ion}}$ which were obtained from the curves in
Fig.~\ref{ions}. While the ionization and detection curves show a
moderate dependence on these parameters, the temperature distribution
given in Fig.~\ref{tempdistr} is very robust. The reason is that only the
absorption cross section $\sigma\left( \mathrm{T}_{1}\right)$ enters
in this figure, which is much better specified by the fitting procedure
than $A_{\text{ion}}$. Incidentally, only this
robust temperature distribution enters the calculation in the
following treatment of radiative decoherence.

\section{Decoherence by thermally emitted radiation }
\label{sec:decoherence} 

The effect of the molecular temperature on the wave nature of the
fullerenes can now be measured using the Talbot-Lau interference
effect described in Sect.~\ref{sec:interferometer}. The  ``fast''
fullerenes with $v=190$\,m/s correspond to the first order Talbot
effect. They are heated by an arrangement of 16 laser beams. Ten beams
are used for the slower fullerenes that are required for the
second Talbot order ($v=100$\,m/s).

\begin{figure}
[ptb]
\includegraphics[width=\columnwidth]
{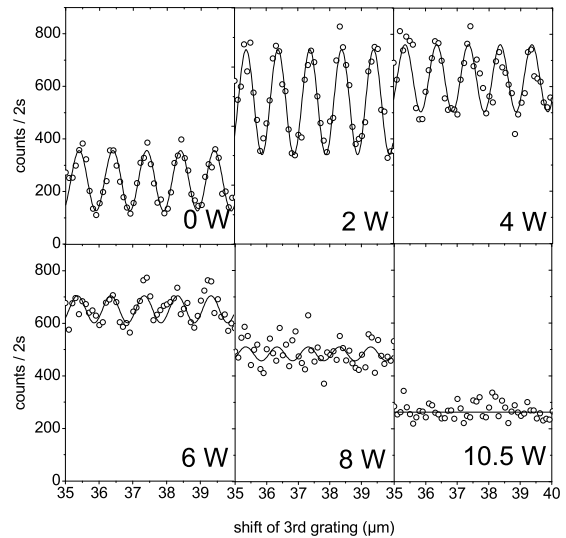}
\caption{Fullerene interference fringes corresponding to the
  first-order Talbot-Lau effect at different powers of the heating
  laser.  The stronger the heating in front of the interferometer the
  more thermal radiation is emitted within the interferometer, leading
  to a monotonic reduction of fringe contrast.  The variation of the
  mean absolute count rate is caused by the thermal ionization
  detection scheme.  While it is irrelevant for the assessment of the
  degree of coherence, it provides us with a second method for the
  evaluation of the molecular temperature, as discussed in the text
  and indicated in Fig.~\ref{detectorcurves}.  The solid curves are
  sinusoidal fits used to extract the fringe visibility.  }
\label{heatedfringes}
\end{figure}

Figure \ref{heatedfringes} shows the observed interference fringes of
the fast molecules for various heating powers. The variation of the
mean count rate is discussed in Sect.~\ref{sec:temperatures}. We use
sinusoidal fits (the solid lines) in order to extract the visibility
$V=(I_{\rm max}-I_{\rm min})/(I_{\rm max}+I_{\rm min})$, which is a
direct measure of the particle's wave nature.

According to decoherence theory, the ability of a hot object to show
interference is limited by the rate and wavelengths of the thermally
emitted photons.  A detailed description of the theoretical treatment
of decoherence in the Talbot-Lau interference has been given in
Ref.~\cite{Hornberger2004a}. It predicts the reduction of the fringe
contrast due to the emission of thermal photons, in particular for the
sinusoidal fringes observed in our symmetric Talbot-Lau
interferometer. For a molecule with the initial temperature $T_{0}$
and a velocity $v$ the reduction factor is given by
\begin{widetext}
\begin{equation}
\mathcal{R}\left(T_{0},v\right)  =\exp\left(  -\int_{0}^{2L/v_{}}
\mathrm{d}t\int_{0}^{\infty}\mathrm{d}\omega~R_{\omega}(\omega,T\left(
t;T_{0}\right)  )\left[  1-\operatorname{sinc}\left(  \frac{\omega d}{c}
\frac{L-|vt-L|}{{L_{\mathrm{T}}}}\right)  \right]  \right)  . \label{Vred}
\end{equation}
\end{widetext}
Here, $R_{\omega}\left( \omega,T\right) $ is the spectral photon
emission rate (\ref{Romega}) at angular frequency $\omega$ and
temperature $T$. The function $T\left( t;T_{0}\right) $ gives the
temporal variation of the fullerene temperature. The argument of the
function $\operatorname{sinc}\left( x\right) =\sin\left( x\right) /x$
compares the wavelength $\lambda=2\pi c/\omega$ of the thermal photon
to the effective separation of the contributing interfering paths.

Equation (\ref{Vred}) is most intuitive for first (second) order
Talbot interference where the grating distance $L$ equals (equals
twice) the Talbot length $L_{\rm T}={d^2}/{\lambda_{\rm C70}}$. Then, the photon
wavelength gets compared to the actual separation of paths running
through neighboring slits (next to neighboring slits) in the central
grating.

It should be noted that the change of the {\em internal} fullerene state
upon photon-emission does not matter for decoherence. The reason for
this is that the internal degrees of freedom do not get entangled with
the center-of-mass coordinate, since the emission probability is
position independent.

The precise form of the spectral photon emission rate $R_{\omega}$ for
C$_{70}$ fullerenes and their temperature evolution were discussed in
Sect.~\ref{sec:cooling}.  In order to predict the expected loss of
interference contrast one needs to account for the temperature
distribution shown by the fullerenes when they enter the
interferometer.  It is given by
\begin{equation}
f_{\mathrm{G1}}(T;v)=\int\mathrm{d}y~i_{\mathrm{C70}}\left(
T,y;v,z_{\mathrm{G1}}\right)  
\end{equation}
since the vertical position is irrelevant in the interferometer. The
expected visibility reduction is calculated by weighting the reduction
factor (\ref{Vred}) with this distribution.
\begin{equation}
\mathcal{R}\left( v\right)
\mathcal{=}\int_0^\infty\mathrm{d}T_{0}~f_{\mathrm{G1}
}(T_{0};v)~\mathcal{R}\left(T_{0},v\right) 
\label{Rmean}
\end{equation}
We use a numeric integration of (\ref{Tdeq}) in order to evaluate the
required function $T\left( t;T_{0}\right)$.

It turns out that the cooling within the interferometer contributes to
the quantitative prediction of the interference contrast. The
visibility loss due to emissions in the first half of the symmetric
interferometer is in general much stronger than in the second half.
At the same time, Eq.~(\ref{Romega}) predicts that unheated molecules,
thermalized only to the oven temperature of 900\,K, do not cool or
decohere noticeably during their time of flight in the apparatus.

As discussed in Sect.~\ref{sec:cooling} the spectral emission rate of
fullerenes deviates considerably from the Boltzmann law of a black
body.  This is one of the reasons why the simple estimates on
temperature effects in Refs.~\cite{Joos1985a,Tegmark1993a,Alicki2002a}
can not be used to describe the experiment.  Similarly, the more
recent studies in Refs.~\cite{Facchi2002a,Viale2003a} do not apply to
the present experimental situation.

\begin{figure*}
[ptb]
\includegraphics[width=0.49\textwidth]{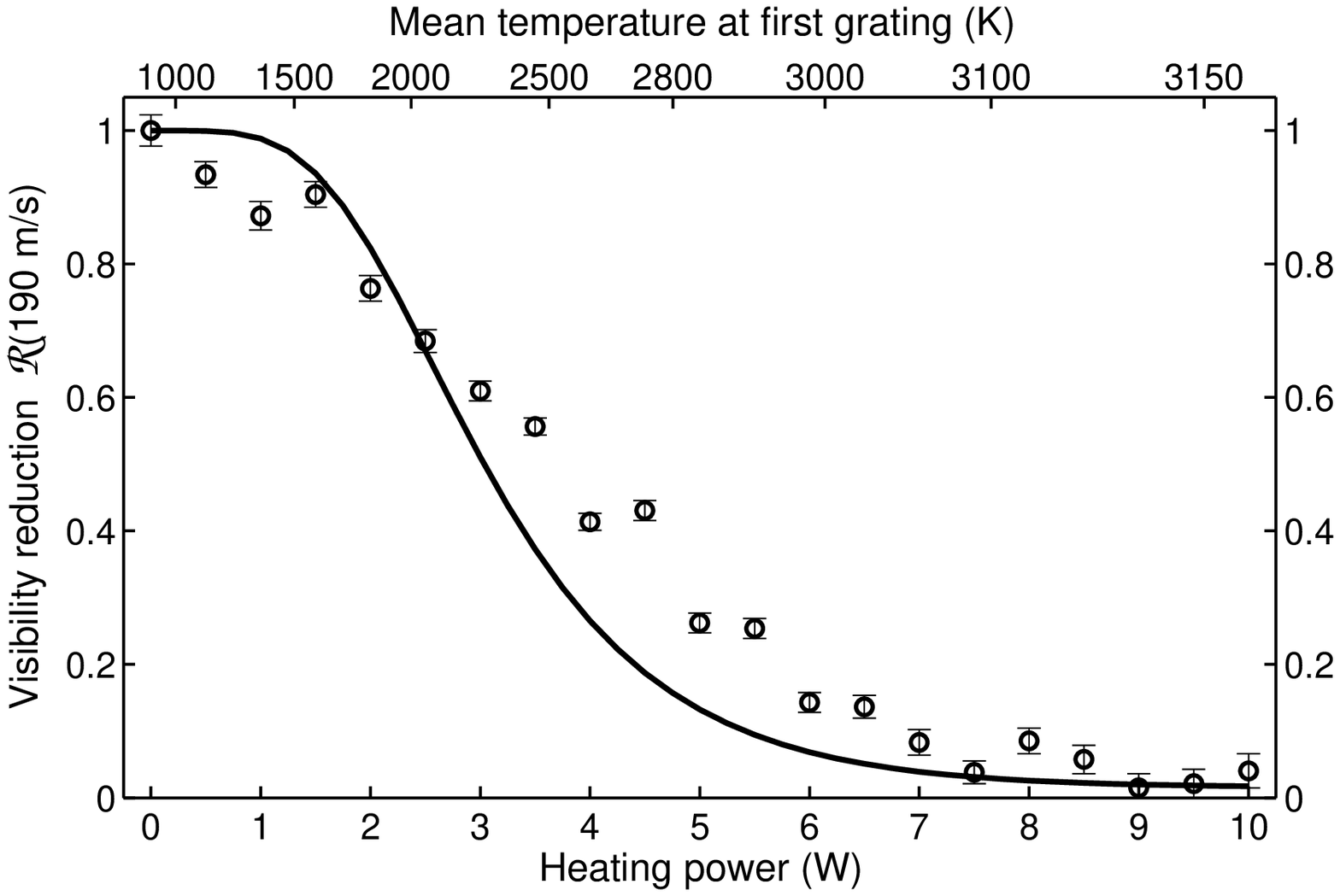}
\includegraphics[width=0.49\textwidth]{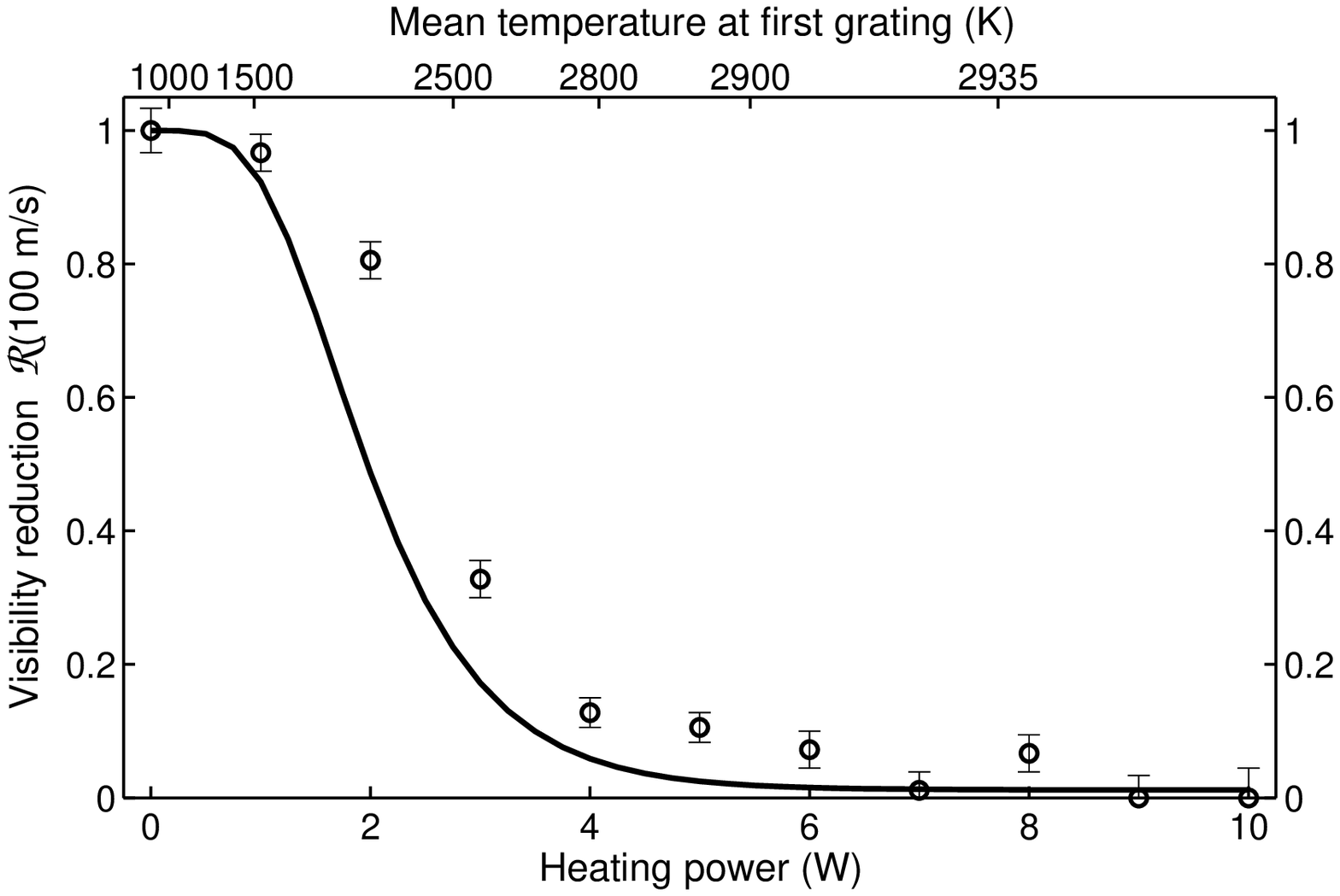}
\caption{Reduction of the interference visibility due to heating with
  laser power $P$. The circles give the experimental visibility
  extracted from fringe patterns such as shown in
  Fig.~\ref{heatedfringes}; the solid lines are the theoretical
  expectation given by Eq.~(\ref{Rmean}). The fullerenes in the left
  graph have a velocity of 190 m/s corresponding to the first Talbot
  order, while in the right graph the fullerene velocity is 100 m/s
  corresponding to the second Talbot order. The upper scale gives the
  mean fullerene temperature at the entrance of the interferometer.}
\label{visibreduction}
\end{figure*}

The visibilities extracted from the experimental fringe patterns are
given in Fig.~\ref{visibreduction}, where they are compared to the
theoretical expectation (\ref{Rmean}). The error bars of the
experimental data indicate only the statistical error of the
visibility extraction. A much larger systematic error is due to the
imperfections in the alignment of the heating laser beams. As a
result, the variation of the experimental data after different
alignments is of the order of the difference between the data shown
and the experimental curve.  Note that the upper scale in
Fig.~\ref{visibreduction} gives the mean temperature in the beam at
the entrance of the interferometer.

\section{Discussion}
\label{sec:conclusions}

The good agreement between experiment and decoherence theory found in
Fig.~\ref{visibreduction} provides strong evidence for the observation
of thermal decoherence. However, since the temperature measurement is
indirect it is necessary to discuss the possibility of other
laser-induced decoherence mechanisms that might give rise to a similar
observation.

A different source of decoherence would ensue if the fullerenes were
still in the electronic triplet state when they enter the
interferometer. This state is associated with a magnetic moment and
arguably also with a higher electric polarizability. The molecules
could therefore be slightly more susceptible to interactions with
electro-magnetic stray fields.

However, if this interaction were relevant one would expect a
different relation between the interference fringe contrast and the
heating laser power than it is observed in the experiment. A single
absorbed photon would suffice to take the fullerene into the first
excited triplet state with a probability of more than 90\%.
Correspondingly, we would expect a sharp threshold in the contrast
curve of Fig.~\ref{visibreduction} when the laser power is increased.
Moreover, we expect that the T$_{1}$ lifetime and therefore also its
population in the interferometer should decrease with increasing
heating power.

From the gradual decay of the visibility curve we can therefore
safely exclude any effect that requires a nonthermal population
of the excited state for the explanation of the observation.

\section*{Acknowledgements}
We acknowledge contributions by Björn Brezger at an early stage of
the work, and we thank Anton Zeilinger for his continuous and helpful
support. This work was supported by the Austrian FWF in the
programs START Y177 and SFB F1505, by the European Union within
the project HPRN-CT-2002-00309 and by the Emmy-Noether program of
the Deutsche Forschungsgemeinschaft.


\begin{thebibliography}{10}

\bibitem{Alicki2002a}
R.~Alicki, \emph{Search for a border between classical and quantum worlds},
  Phys. Rev. A \textbf{65} (2002), 034104.

\bibitem{Andersen2000a}
J.~U. Andersen and E.~Bonderup, \emph{Classical dielectric models of fullerenes
  and estimation of heat radiation}, Eur. Phys. J. D \textbf{11} (2000), 413 --
  434.

\bibitem{Arndt1999a}
M.~Arndt, O.~Nairz, J.~Voss-Andreae, C.~Keller, G.~Van der Zouw, and
  A.~Zeilinger, \emph{Wave-particle duality of {C}$_{60}$ molecules}, Nature
  \textbf{401} (1999), 680--682.

\bibitem{Bertet2001a}
P.~Bertet, S.~Osnaghi, A.~Rauschenbeutel, G.~Nogues, A.~Auffeves, M.~Brune,
  J.~M. Raimond, and S.~Haroche, \emph{A complementarity experiment with an
  interferometer at the quantum-classical boundary}, Nature \textbf{411}
  (2001), 166 -- 170.

\bibitem{Borde1994a}
Ch.J. Bord{\'e}, N.~Courtier, F.~Du Burck, A.N. Goncharov, and M.~Gorlicki,
  \emph{Molecular interferometry experiments}, Phys. Lett. A \textbf{188}
  (1994), 187--197.

\bibitem{Brezger2002a}
B.~Brezger, L.~Hackerm{\"u}ller, S.~Uttenthaler, J.~Petschinka, M.~Arndt, and
  A.~Zeilinger, \emph{Matter-wave interferometer for large molecules}, Phys.
  Rev. Lett. \textbf{88} (2002), 100404.

\bibitem{Brune1996a}
M.~Brune, E.~Hagley, J.~Dreyer, X.~Ma{\^{\i}}tre, A.~Maali, C.~Wunderlich,
  J.~M. Raimond, and S.~Haroche, \emph{Observing the progressive decoherence of
  the ``meter'' in a quantum measurement}, Phys. Rev. Lett. \textbf{77} (1996),
  4887 -- 4890.

\bibitem{Chapman1995a}
M.~S. Chapman, T.~D. Hammond, A.~Lenef, J.~Schmiedmayer, R.~A. Rubenstein,
  E.~Smith, and D.~E. Pritchard, \emph{Photon scattering from atoms in an atom
  interferometer: {C}oherence lost and regained}, Phys. Rev. Lett. \textbf{75}
  (1995), 3783 -- 3787.

\bibitem{Chapman1995b}
Michael~S. Chapman, Christopher~R. Ekstrom, Troy~D. Hammond, Richard~A.
  Rubenstein, J{\"o}rg Schmiedmayer, Stefan Wehinger, and David~E. Pritchard,
  \emph{Optics and interferometry with {Na}$_2$ molecules}, Phys. Rev. Lett.
  \textbf{74} (1995), 4783--4786.

\bibitem{Clauser1994a}
John~F. Clauser and Shifang Li, \emph{Talbot-{vonLau} atom interferometry with
  cold slow potassium}, Phys. Rev. A \textbf{49} (1994), R2213.

\bibitem{Coheur1996a}
P.~F. Coheur, M.~Carleer, and R.~Colin, \emph{The absorption cross sections of
  {C}$_{60}$ and {C}$_{70}$ in the visible-{UV} region}, J. Phys. B: At. Mol.
  Opt. Phys. \textbf{29} (1996), 4987 -- 4995.

\bibitem{Ding1994a}
D.~Ding, J.~Huang, R.~N. Compton, C.~E. Klots, and R.~E. Haufler, \emph{cw
  laser ionization of {C}$_{60}$ and {C}$_{70}$}, Phys. Rev. Lett. \textbf{73}
  (1994), no.~8, 1084 -- 1087.

\bibitem{Dresselhaus1998a}
M.~S. Dresselhaus, G.~Dresselhaus, and P.~C. Eklund, \emph{Science of
  fullerenes and carbon nanotubes}, 2 ed., Acad. Press, San Diego, 1998.

\bibitem{Facchi2002a}
P.~Facchi, A.~Mariano, and S.~Pascazio, \emph{Mesoscopic interference}, Recent
  Res. Devel. Physics \textbf{3} (2002), 1--29.

\bibitem{Friedrich1998a}
H.~Friedrich, \emph{Theoretical atomic physics}, Springer, Berlin, 1998.

\bibitem{Hackermuller2004a}
L.~Hackerm{\"u}ller, K.~Hornberger, B.~Brezger, A.~Zeilinger, and M.~Arndt,
  \emph{Decoherence of matter waves by thermal emission of radiation}, Nature
  \textbf{427} (2004), 711--714.

\bibitem{Hackermuller2003a}
L.~Hackerm{\"u}ller, S.~Uttenthaler, K.~Hornberger, E.~Reiger, B.~Brezger,
  A.~Zeilinger, and M.~Arndt, \emph{Wave nature of biomolecules and
  fluorofullerenes}, Phys. Rev. Lett. \textbf{91} (2003), 90408.

\bibitem{Hansen1998a}
K.~Hansen and E.~E.~B. Campbell, \emph{Thermal radiation from small particles},
  Phys. Rev.~E \textbf{58} (1998), 5477.

\bibitem{Heszler1997a}
P.~Heszler, J.~O. Carlsson, and P.~Demirev, \emph{Photon emission from gas
  phase fullerenes excited by 193 nm laser radiation}, Chem. Phys. \textbf{107}
  (1997), 10440--10445.

\bibitem{Hornberger2004a}
K.~Hornberger, J.~E. Sipe, and M.~Arndt, \emph{Theory of decoherence in a
  matter mave {T}albot-{L}au interferometer}, Phys. Rev. A \textbf{70} (2004),
  053608.

\bibitem{Ji2003a}
Y.~Ji, Y.~Chung, D.~Sprinzak, M.~Heiblum, D.~Mahalu, and H.~Shtrikman, \emph{An
  electronic mach-zehnder interferometer}, Nature \textbf{422} (2003), 415 --
  418.

\bibitem{Joos1985a}
E.~Joos and H.~D. Zeh, \emph{The emergence of classical properties through
  interaction with the environment}, Z. Phys. B. \textbf{59} (1985), 223--243.

\bibitem{Joos2003a}
E.~Joos, H.~D. Zeh, C.~Kiefer, D.~Giulini, J.~Kupsch, and I.-O. Stamatescu,
  \emph{Decoherence and the appearance of a classical world in quantum theory},
  2nd ed., Springer, Berlin, 2003.

\bibitem{Klots1991a}
C.~E. Klots, \emph{Quasiequilibrium rate constants for thermionic emission from
  small particles}, Chem. Phys. Lett. \textbf{186} (1991), 73 -- 76.

\bibitem{Kokorowski2001a}
D.~A. Kokorowski, A.~D. Cronin, T.~D. Roberts, and D~.E. Pritchard, \emph{From
  single- to multiple-photon decoherence in an atom interferometer}, Phys. Rev.
  Lett. \textbf{86} (2001), 2191.

\bibitem{Kolodney1995a}
E.~Kolodney, B.~Tsipinyuk, and A.~Budrevich, \emph{The thermal energy
  dependence (10--20 {eV)} of electron impact induced fragmentation of C60 in
  molecular beams: Experiment and model calculations}, J. Chem. Phys.
  \textbf{102} (1995), 9263 -- 9275.

\bibitem{Matt1999a}
S.~Matt, O.~Echt, M.~Sonderegger, R.~David, P.~Scheier, J.~Laskin, C.~Lifshitz,
  and T.~D. M{\"a}rk, \emph{Kinetic energy release distribution and evaporation
  energies for metastable fullerene ions}, Chem. Phys. Lett. \textbf{303}
  (1999), 379 -- 386.

\bibitem{Mitzner1995a}
R.~Mitzner and E.~E.~B. Campbell, \emph{Optical emission studies of laser
  desorbed {C}$_{60}$}, J. Chem. Phys. \textbf{103} (1995), 2445--2453.

\bibitem{Nairz2000a}
O.~Nairz, M.~Arndt, and A.~Zeilinger, \emph{Experimental challenges in
  fullerene interferometry}, J. Mod. Opt. \textbf{47} (2000), 2811--2821.

\bibitem{Nairz2001a}
O.~Nairz, B.~Brezger, M.~Arndt, and A.~Zeilinger, \emph{Diffraction of complex
  molecules by structures made of light}, Phys. Rev. Lett. \textbf{87} (2001),
  160401--1 -- 160401--4.

\bibitem{Reinkoster2001a}
A.~Reink{\"o}ster, U.~Werner, N.~M. Kabachnik, and H.~O. Lutz,
  \emph{Experimental and theoretical study of ionization and fragmentation of
  C-60 by fast-proton impact}, Phys. Rev. A \textbf{64} (2001), 023201.

\bibitem{Rohlfing1988a}
E.~A. Rohlfing, \emph{Optical emission studies of atomic, molecular, and
  particulate carbon produced from a laser vaporization cluster source}, J.
  Chem. Phys \textbf{89} (1988), 6103--6112.

\bibitem{Schoellkopf2004a}
W.~Schoellkopf, R.E. Grisenti, and J.P. Toennies, \emph{Time-of-flight resolved
  transmission-grating diffraction of molecular beams}, Eur. Phys. J. D
  \textbf{28} (2004), 125 -- 133.

\bibitem{Schollkopf1996a}
W.~Sch{\"o}llkopf and J.P. Toennies, \emph{The nondestructive detection of the
  helium dimer and trimer}, J. Chem. Phys. \textbf{104} (1996), 1155--1158.

\bibitem{Scoles1988a}
G.~Scoles, D.~Bassi, U.~Buck, and D.~Lain{\'e} (eds.), \emph{Atomic and
  molecular beam methods}, vol.~I, Oxford University Press, 1988.

\bibitem{Tegmark1993a}
M.~Tegmark, \emph{Apparent wave function collapse caused by scattering}, Found.
  Phys. Lett. \textbf{6} (1993), 571--590.

\bibitem{Viale2003a}
A.~Viale, M.~Vicari, and N.~Zanghi, \emph{Analysis of the loss of coherence in
  interferometry with macromolecules}, Phys. Rev. A \textbf{68} (2003), 063610.

\end{thebibliography}


\end{document}